\documentclass[aps,twocolumn,prl,superscriptaddress,longbibliography]{revtex4-2}
\usepackage{graphicx}% Include figure files
\usepackage{dcolumn}% Align table columns on decimal point
\usepackage{bm}% bold math
\usepackage[T1]{fontenc}
\usepackage{textcomp}
\usepackage{amsmath}
\usepackage{mathrsfs}
\usepackage{amssymb}
\usepackage{units}
\usepackage{ulem}
\usepackage{xcolor}
\usepackage[colorlinks,bookmarks=false,citecolor=blue,linkcolor=blue,urlcolor=blue]{hyperref}
\usepackage{soul}

\begin{document}

\title{Excitonic oscillator-strength saturation dominates polariton-polariton interactions}

\author{Maxime Richard}
\affiliation{Majulab International Research Laboratory, French National Centre for Scientifique Research, National University of Singapore, Nanyang Technological University, Sorbonne Universit\'e, Universit\'e C\^ote d'Azur, 117543 Singapore}
\affiliation{Centre for Quantum technologies, National University of Singapore, 117543 Singapore}

\author{Ir\'en\'ee Fr\'erot}
\affiliation{Laboratoire Kastler Brossel, Sorbonne Universit\'e, CNRS, ENS-PSL Research University, Coll\`ege de France, 4 Place Jussieu, 75005 Paris, France}

\author{Sylvain Ravets}
\affiliation{Centre de Nanosciences et de Nanotechnologies, CNRS, Universit\'{e} Paris-Sud, Universit\'{e} Paris-Saclay, 91120 Palaiseau, France}

\author{Jacqueline Bloch}
\affiliation{Centre de Nanosciences et de Nanotechnologies, CNRS, Universit\'{e} Paris-Sud, Universit\'{e} Paris-Saclay, 91120 Palaiseau, France}

\author{Carlos Anton-Solanas}
\affiliation{Departamento de F\'isica de Materiales, Instituto Nicol\'as Cabrera, Instituto de F\'isica de la Materia Condensada, Universidad Aut\'onoma de Madrid, 28049 Madrid, Spain}

\author{Ferdinand Claude}
\affiliation{Majulab International Research Laboratory, French National Centre for Scientifique Research, National University of Singapore, Nanyang Technological University, Sorbonne Universit\'e, Universit\'e C\^ote d'Azur, 117543 Singapore}
\affiliation{Centre for Quantum technologies, National University of Singapore, 117543 Singapore}

\author{Yueguang Zhou}
\affiliation{Majulab International Research Laboratory, French National Centre for Scientifique Research, National University of Singapore, Nanyang Technological University, Sorbonne Universit\'e, Universit\'e C\^ote d'Azur, 117543 Singapore}
\affiliation{Centre for Quantum technologies, National University of Singapore, 117543 Singapore}

\author{Martina Morassi}
\affiliation{Centre de Nanosciences et de Nanotechnologies, CNRS, Universit\'{e} Paris-Sud, Universit\'{e} Paris-Saclay, 91120 Palaiseau, France}

\author{Aristide Lema\^itre}
\affiliation{Centre de Nanosciences et de Nanotechnologies, CNRS, Universit\'{e} Paris-Sud, Universit\'{e} Paris-Saclay, 91120 Palaiseau, France}

\author{Iacopo Carusotto}
\affiliation{Pitaevskii BEC Center, INO-CNR and Dipartimento di Fisica, Universit\`{a} di Trento, via Sommarive 14, I-38123 Trento, Italy}

\author{Anna Minguzzi}
\affiliation{Univ. Grenoble Alpes, CNRS, LPMMC, 38000 Grenoble, France}

%\date{\today}

\begin{abstract}
Exciton-polaritons in semiconductor microcavities exhibit large two-body interactions that, thanks to ever refined nanotechnology techniques, are getting closer and closer to the quantum regime where single-photon nonlinearities start being relevant. To foster additional progress in this direction, in this work we experimentally investigate the microscopic mechanism driving polariton-polariton interactions. We measure the dispersion relation of the collective excitations that are thermally generated on top of a coherent fluid of interacting lower-polaritons. By comparing the measurements with the Bogoliubov theory over both the lower and upper polariton branches simultaneously, we find that polariton-polariton interactions stem dominantly from a mechanism of saturation of the exciton oscillator strength. 

\end{abstract}

\maketitle

\begin{figure*}[t]
   \includegraphics[width=0.85\textwidth]{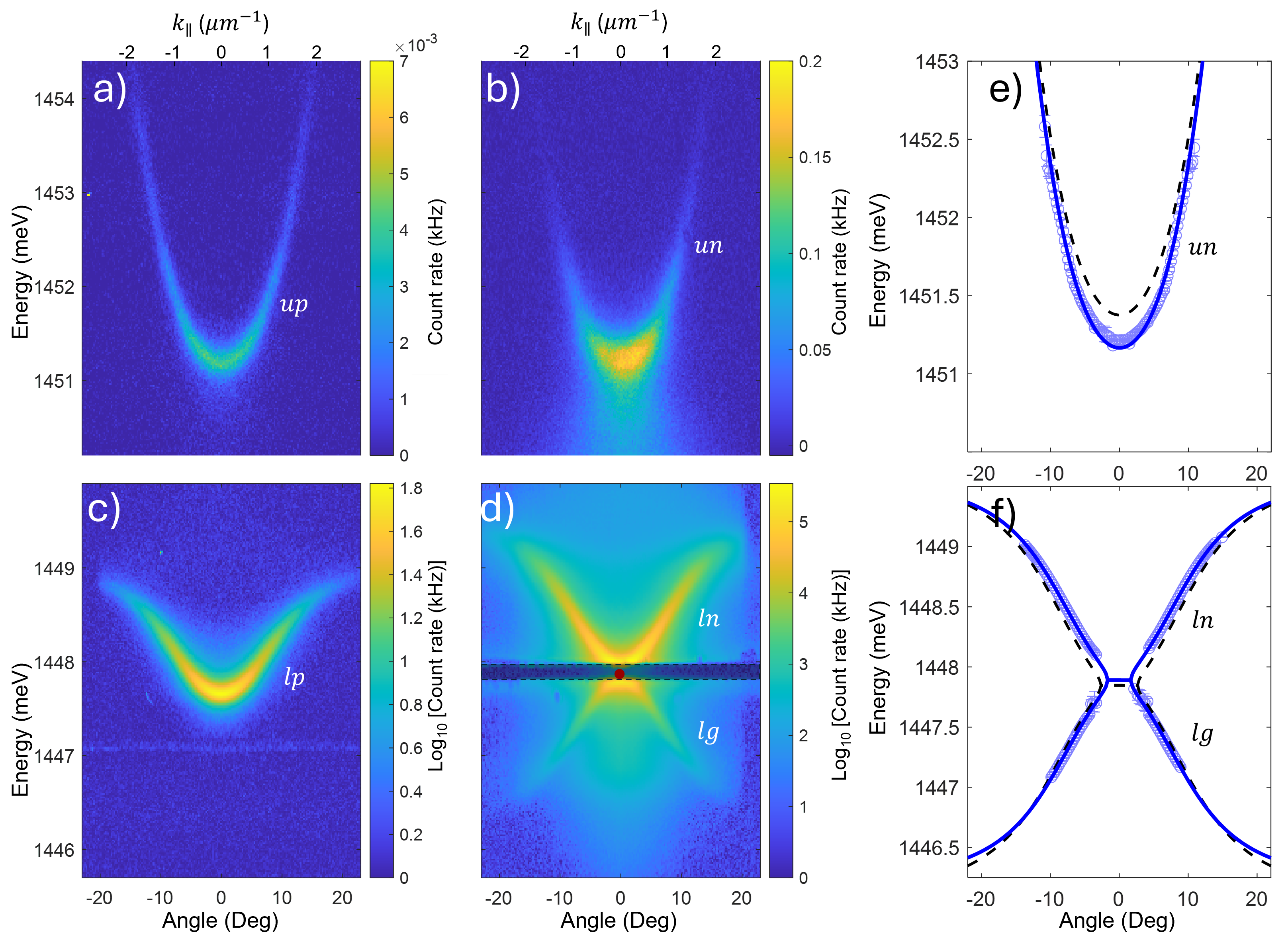}
\caption{Analysis of working point WP3c (see main text) - Left panels: Photoluminescence maps $I_{PL}(\theta,\omega)$ of (a) the upper- and (c) the lower-polariton branches obtained under weak non-resonant excitation (with vanishing two-body interactions). Central panels: emission map $I_{B}(\theta,\omega)$ of (b) the upper-polariton normal (\textit{un}), and (d) lower-polariton normal (\textit{ln}) and ghost (\textit{lg}) (negative curvature) Bogoliubov excitations branches under resonant excitation. The dashed rectangle in (d) shows where the mean field emission is rejected by the spectral filter, and the red circle indicates $\omega_{\rm las}$. Right panels: Measured and fitted dispersion relation of (e) the \textit{un} and (f) the \textit{ln} and \textit{lg} branches. The blue symbols are the measurements obtained from analysis of the data in panels (b) and (d). The solid blue (dashed black) line is a fit using Bogoliubov theory, and leaving $g_{xx}n_x$ and $g_{xc}n_x$ as free parameters (leaving $g_{xx}n_x$ as free parameter, and setting $g_{xc}n_x=0$). See main text for details.}
\label{fig1}
\end{figure*}

% INTRODUCTION
%interacting photons [define: hbar g_0, gamma]
In the broad context of fundamental and applied quantum sciences, visible photons constitute versatile and powerful quantum degrees of freedom \cite{OBrien_2009,Pelucchi_2022,Luo_2023}. Quantum states of light can indeed be prepared and sent over long distances with minimal losses in free space \cite{Vallone_2015,Liao_2017,Liao2_2017} or in optical-fiber networks \cite{Marcikic2003,Korzh_2015,Paraiso_2021} for quantum communication, and manipulated in advanced photonic circuits to perform quantum computing tasks \cite{Wang_2020,Madsen_2022,Maring_2024}. These capabilities are enabled by the fact that visible photons can be detected individually with a high quantum efficiency, and by the fact that they can be spatially and spectrally manipulated by engineering the dielectric environment \cite{Silverstone_2016,Schneider_2016,Senellart_2017}. But unlike other quantum systems such as superconducting circuits, trapped ions, or ultra-cold atoms, making visible photons to strongly interact with each other at the single-particle level~\cite{Chang_2014} remains a daunting challenge in practical and scalable solid-state systems.

%polaritons as interacting photons [define: g=Ag_0; \cal F=g_0\gamma]
In recent years, significant progress in this direction has been made by 
engineering effective interactions between photons propagating in solid-state systems, in order to dress photons with a suitable electronic transition from, e.g., a quantum dot~\cite{Loo_2012,Javadi_2015,Lodahl_2015,DeSantis_2017} or a quantum well. ~\cite{Munoz_2019,Delteil_2019,Scarpelli_2024,Kuriakose_2022}. In the latter case, when the quantum well is embedded in a microcavity, the system can operate in the strong coupling regime between excitons and photons leading to exciton-polaritons (in short, polaritons). The dressing thus exceeds the perturbative regime and the resulting exciton-mediated two-body interaction is dramatically enhanced. 

In GaAs-based open-microcavity systems \cite{Besga_2015}, a polariton-polariton interaction figure of merit ${\cal F}=U/\gamma \simeq 0.15$ has been reported \cite{Munoz_2019,Delteil_2019}, where $\hbar U$ is the interaction strength between two polaritons, and $\hbar\gamma$ is the polariton spectral linewidth. In order to enter the quantum regime ${\cal F}\gtrsim 1$, different strategies are currently studied, such as adding a static dipole moment to excitons to increase Coulomb interactions \cite{Togan_2018,Rosenberg_2018,Datta_2022,Liran_2024}, coupling polaritons to a sea of free charges to obtain bulkier polaron-polaritons~\cite{Libing_2020} or quantum Hall polaritons~\cite{Knuppel_2019}, using large Rydberg excitons with high principal quantum number \cite{Kaz_2014}, or exploiting biexcitonic Feshbach resonances in the two-polariton scattering~\cite{Wouters2_2007,Carusotto_2010,Takemura_2014,Scarpelli_2024}. A major difficulty along this path is that a precise understanding of the mechanism dominating polariton-polariton interactions is still lacking, even in the simplest configurations.

Indeed, while exciton-exciton scattering seems a reasonable mechanism underlying polariton-polariton interactions, first hints that a saturation mechanism of the excitonic oscillator strength might contribute significantly as well, were recently mentioned in GaAs-based \cite{Frerot_2023} and in TMD-monolayer microcavities \cite{Stepanov_2021,Datta_2022,Zhao_2023}. In order to understand the generality of this unexpected feature, here we develop an accurate and dedicated method to quantitatively discriminate between saturation mechanism and exciton-exciton scattering. We investigate a typical GaAs-based microcavity over a large range of exciton-photon detuning and of interaction energy, and we measure the dispersion relation of the collective Bogoliubov excitations of the coherent polariton fluid~\cite{Carusotto_2013}. We demonstrate that they contain 
an accurate and accessible information on the mechanism governing polariton-polariton interactions, especially when upper- and lower-polariton excitations are simultaneously considered. For all investigated parameters, we thus find that polariton-polariton interaction is consistently dominated by a saturation mechanism of the excitonic oscillator strength, and we rule out other plausible explanations.

% theoretical description of interacting polaritons 
% define: \Omega, \omega_x; \omega_{c,k}; k, n_cav; \hat {\cal H}_{xx}; \hat {\cal H}_{xc}; g_xx; g_xc; \hat{M}_B(), {lg,ln,ug,un}
In order to model interacting polaritons in this context, we describe the system of strongly coupled excitons and photons confined in a planar microcavity by the Hamiltonian 
\begin{eqnarray}
\hat{\cal H} & = & \hbar \sum_{\bf k} \left[ \omega_x \hat b_{\bf k}^\dagger \hat b_{\bf k} + \omega_{c,{\bf k}} \hat a_{{\bf k}}^\dagger \hat a_{{\bf k}} + \frac{\Omega}{2} (\hat a_{\bf k}^\dagger \hat b_{\bf k} + {\rm h.c.})\right]+ \nonumber \\ 
&& + {\hat F} + \hat {\cal H}_{xx} + \hat {\cal H}_{xc},  \label{eq:Ham-sys} 
\end{eqnarray}
where ${\hat b}_{\bf k}^\dagger$ and ${\hat a}_{\bf k}^\dagger$ are the bosonic creation operators for  excitons and photons respectively,  with in-plane momentum $\hbar\bf k$, coupled to each other at a Rabi frequency $\Omega/2$, and driven resonantly by a coherent field of in-plane momentum $\hbar\bf k_p$ coupled to cavity photons as ${\hat F}=({\hat a}_{\bf k_p}F^*_0e^{i\omega_{\rm las}t} + {\rm h.c.})$. The excitonic transition frequency $\omega_x$ can be safely considered constant inside the polariton-relevant light cone, and the cavity-mode dispersion relation has the form $\omega_{c,{\bf k}}^2=\omega_{c,0}^2+c^2k^2/n_{cav}^2$, where $n_{cav}$ is the effective refractive index of the cavity mode. 

As a main focus of this work, the interaction terms $\hat {\cal H}_{xx}$ and $\hat {\cal H}_{xc}$ model the exciton-exciton scattering and excitonic-saturation-mediated interaction terms. In detail,
\begin{equation}
\hat {\cal H}_{xx} = \frac{\hbar g_{xx}}{2} \sum_{{\bf k},{\bf k}',{\bf q}}
\hat b_{{\bf k}+{\bf q}}^\dagger \hat b_{{\bf k}'-{\bf q}}^\dagger \hat b_{{\bf k}'} \hat b_{\bf k}
\label{H_xx}
\end{equation}
describes exciton-exciton scattering with coupling strength $\hbar g_{xx}$ \footnote{The different interaction strengths $\hbar g$ used in this work are defined for a 2D polaritonic system; as a result, they are related to the two-polaritons interaction strength $U$ as $g\propto U\cal{A}$, where $\cal{A}$ is the in-plane polariton mode area.}. In \cite{Ciuti_1998}, $\hbar g_{xx}$ was determined to be approximately constant in momentum space within the light cone, and with a magnitude determined by the carrier-exchange contribution to Coulomb interaction. 

When the exciton-photon coupling is included in the model, either in the weak~\cite{Rochat_2000} or in the strong coupling regime relevant for polaritons \cite{Tassone_1999,Glazov_2009}, another interaction mechanism is found which is rooted into the fermionic nature of the electrons and holes constituting excitons, which leads to~\cite{Rochat_2000,Tassone_1999,Glazov_2009}
\begin{equation}
\hat {\cal H}_{xc} = -\frac{\hbar g_{xc}}{2}\sum_{{\bf k},{\bf k}',{\bf q}} (\hat a_{{\bf k}+{\bf q}}^\dagger \hat b_{{\bf k}'-{\bf q}}^\dagger \hat b_{{\bf k}'} \hat b_{\bf k} + {\rm h.c.}),
\label{H_xc}
\end{equation}
where $\hbar g_{xc}$ is the interaction strength. The main effect of this term is to reduce the effective exciton-photon coupling [of nominal strength $\hbar\Omega$ in Eq.~\eqref{eq:Ham-sys}] as the excitonic density $n_x$ increases. 

Owing to the relatively low polariton density involved in the experiment, we can work within a mean-field approximation. Following \cite{Frerot_2023}, the dispersion relation of the Bogoliubov excitations can thus be obtained by expanding to quadratic order the Hamiltonian~(\ref{eq:Ham-sys}) on top of a polaritonic mean-field and then diagonalizing the Bogoliubov matrix $\hat{M}_B({\bf k_p},{\bf k})$ (see SM section II \cite{SuppMat}). The excitation spectrum displays four branches $\omega_{B,j}(k)$: the lower-polaritons ghost (labeled \textit{lg}) and normal (labeled \textit{ln}) branches, and the upper polaritons ghost (labeled \textit{ug}) and normal branches (labeled \textit{un}).

%discussion on the shifts of the Bogoliuv branches
%define: $\omega_{B,j}; omega_{0,lp/up}(k); {lg,ln,ug,un}; \tilde{omega_x}; \tilde{Omega}

%summarized version:
A key characteristic of these branches is the fact that their shape and relative spectral shift depend significantly on the relative contribution of the exciton-exciton (\ref{H_xx}) and saturation (\ref{H_xc}) interaction mechanisms. By examining the structure of $\hat{M}_B$ (and assuming that the laser drives a lower-polariton mean field), one can show that the low Bogoliubov excitations frequency region (bounded by $\pm E_{nl}$ on both sides of the driven mean-field, where $E_{nl}\ll \hbar\Omega$) of the \textit{lg} and and \textit{ln} dispersion relations undergo a change of shape with respect to the non-interacting polaritons dispersion relation, due to the particle-hole coupling (the $2\times 2$ off-diagonal blocks in $\hat{M}_B$, see SM eq.(15) \cite{SuppMat}), which is the hallmark of Bogoliubov excitations~\cite{Carusotto_2013}. At higher frequencies, including in the upper polaritons sector (\textit{ug} and \textit{un}), the Bogoliubov dispersion relations are subject to a simpler spectral shift, which is determined by the $2\times 2$ diagonal blocks in $\hat{M}_B$ (See SM eq.(14) \cite{SuppMat}). Interestingly, this spectral shift of \textit{un}, depends particularly strongly on whether $H_{xx}$ or $H_{xc}$ dominates: When $H_{xx}$ dominates, the effective excitonic transition shifts up by $\tilde{\omega}_x-\omega_x=2g_{xx}n_x$. As a result, both \textit{ln} and \textit{un} shift up by a comparable amount. When $H_{xc}$ dominates, the effective excitonic transition also shifts up, by $\tilde{\omega}_x-\omega_x=2g_{xc}\sqrt{n_xn_c}$, but the effective Rabi splitting is reduced by $\tilde{\Omega}-\Omega=-g_{xc}n_x$. As a result, while these two corrections add up to shift \textit{ln} up, they contribute with an opposite sign to \textit{un}. \textit{un}'s spectral shift is thus much weaker than for the \textit{ln} branch and changes sign around $\delta_{xc}=0$. This difference is a key observable to pinpoint the relative contributions of the $H_{xx}$ and $H_{xc}$ interaction mechanisms in a measurement of $\omega_{B,ln}(k)$, $\omega_{B,un}(k)$.

% Experiment: sample + setup
%define \xi; \R_det; E_nl
In the experiment, we employ a state-of-the-art planar GaAs-based microcavity (identical to that in Ref. \cite{Frerot_2023}) operating in the strong coupling regime between the cavity photons and the excitonic transition of a single high-quality $17\,$nm thick In$_{5\%}$Ga$_{95\%}$As quantum well embedded in the cavity spacer. This spacer is intentionally wedge-shaped, so that the photon-exciton detuning $\delta_{xc}=\hbar(\omega_c(k=0)-\omega_x)$ can be chosen at will by changing the working point a the cavity surface. The sample is operated at $T=3.9\,K$. Polaritons are excited using a continuous wave laser focused into a Gaussian spot of $35\,\mu$m radius, of which only the central part of radius $R_{\rm det}=15\,\mu$m $\gg\xi$ is kept for detection. For the chosen experimental parameters, $\xi\simeq\hbar/\sqrt{2mE_{nl}}<3.5\,\mu$m is the healing length of the lower-polariton quantum fluid, and the characteristic energy scale of the interactions in the polaritonic fluid is varied in the range $E_{nl}=\hbar(g_{xx}+g_{xc})n_x\in[50,300]\,\mu$eV by varying the driving laser intensity. The laser frequency is tuned either (i) far above the polaritonic resonances to excite low-density polariton photoluminescence (Fig.\ref{fig1}.a,c) or (ii) close to resonance with the lowest polariton state at $k_p\simeq 0$ (Fig.\ref{fig1}.b,d). The laser polarization is chosen circular in order to focus on the interaction between co-polarized polaritons. The interaction between cross-polarized polaritons is significantly smaller and involves different physical mechanisms such as biexciton Feshbach enhancement~\cite{Wouters_2007,Carusotto_2010,Takemura_2014,Scarpelli_2024} and intermediate spin-2 exciton states \cite{Vladimirova_2010}: its study goes beyond the scope of this Letter and will be examined in a future work.

% Experiment: determine system parameters (linear) by PL 
%define \delta_xc;  $I_{PL}(\theta,\omega)$
For each investigated working points (WP), we first use the laser setting (i) to collect an angle- and energy-resolved polariton photoluminescence maps $I_{PL}(\theta,\omega)$. The maps obtained for WP3 are shown in Fig.\ref{fig1}.(a,c) The parameters of all 5 investigated WPs can be found in the SM section I \cite{SuppMat}). This allows us to determine the bare polariton dispersion in the non-interacting regime.  for each angle $\theta$, the central position of the low and high energy peaks in $I_{PL}(\theta,\omega)$ provides a measurement of $\hbar\omega_{0,lp}(k)$, and $\hbar\omega_{0,up}(k)$ at $k=\omega\sin(\theta)/c$. The resulting dispersion relations are fitted with a non-interacting polaritonic model described by the bracket term in Eq.~\eqref{eq:Ham-sys} (see SM section I \cite{SuppMat}). Owing to the excellent agreement, this analysis provides a precise measurement of most system parameters. For WP3, we thus obtain $\hbar\Omega=3.49 \pm 0.02\,$meV for the Rabi splitting, $n_{cav}=3.42 \pm 0.04$  for the effective  refractive index of the cavity mode, $\hbar\omega_c(k_\parallel{=}0) = 1449.58\pm0.04\,$meV for the bare cavity resonance energy at $k_\parallel=0$, and $\hbar\omega_x = 1449.25\pm 0.01\,$meV for the excitonic transition energy.

% Experiment: measure Bogo spectral function
%define \delta_las ; $I_B(\theta,\omega)$
We then use the resonant laser setting (ii) where we apply different, slightly blue-detuned laser frequencies $\omega_{\rm las}$ with respect to the non-interacting lower polariton state frequency $\omega_{0,lp}(0)$: $\delta_{\rm las}=\omega_{\rm las}-\omega_{0,lp}(k=0)\in[50,300]\mu$eV. For all investigated WPs, the blue-detuned $\delta_{\rm las}$ is chosen to be large enough for the polaritonic field to be bistable~\cite{Carusotto_2013}. We then drive the system in the high density branch of the bistability, that offers optimal conditions to generate a stable steady-state polaritonic fluid with an interaction energy comparable or larger than $\hbar\delta_{\rm las}$~\cite{Claude_2022,Frerot_2023}. The transmitted coherent light is then rejected by narrow-band spectral filtering, in order to collect only the emission $I_B(\theta,\omega)$ from the Bogoliubov excitations that are dominantly generated by coupling to the thermal phonons of the solid-state environment \cite{Frerot_2023}. The Bogoliubov emission $I_B(\theta,\omega)$ collected in this way at WP3, for $\delta_{\rm las}=203\,\mu$eV and a laser power of $13.7\,$mW, is shown in Fig.\ref{fig1}.(b,d) (experiment labeled "WP3c" thereafter). A list of all investigated experiments "WP$_jk$" can be found in Table (II) of the SM \cite{SuppMat}. for all WPs, these parameters are chosen for the system to be close to the sonic regime of the collective excitations \cite{Carusotto_2013}. For each angle $\theta$, $I_B(\theta,\omega)$ exhibits three resonance peaks corresponding --in increasing energy order-- to the \textit{lg}, \textit{ln} and \textit{un} Bogoliubov states; note that the \textit{ug} resonance yields a too weak emission to be observed in these experimental conditions. The measured dispersion relations extracted from $I_B(\theta,\omega)$ are shown in Fig.\ref{fig1}.(e,f) for WP3c (blue symbols).

\begin{figure}[t]
\includegraphics[width=\columnwidth]{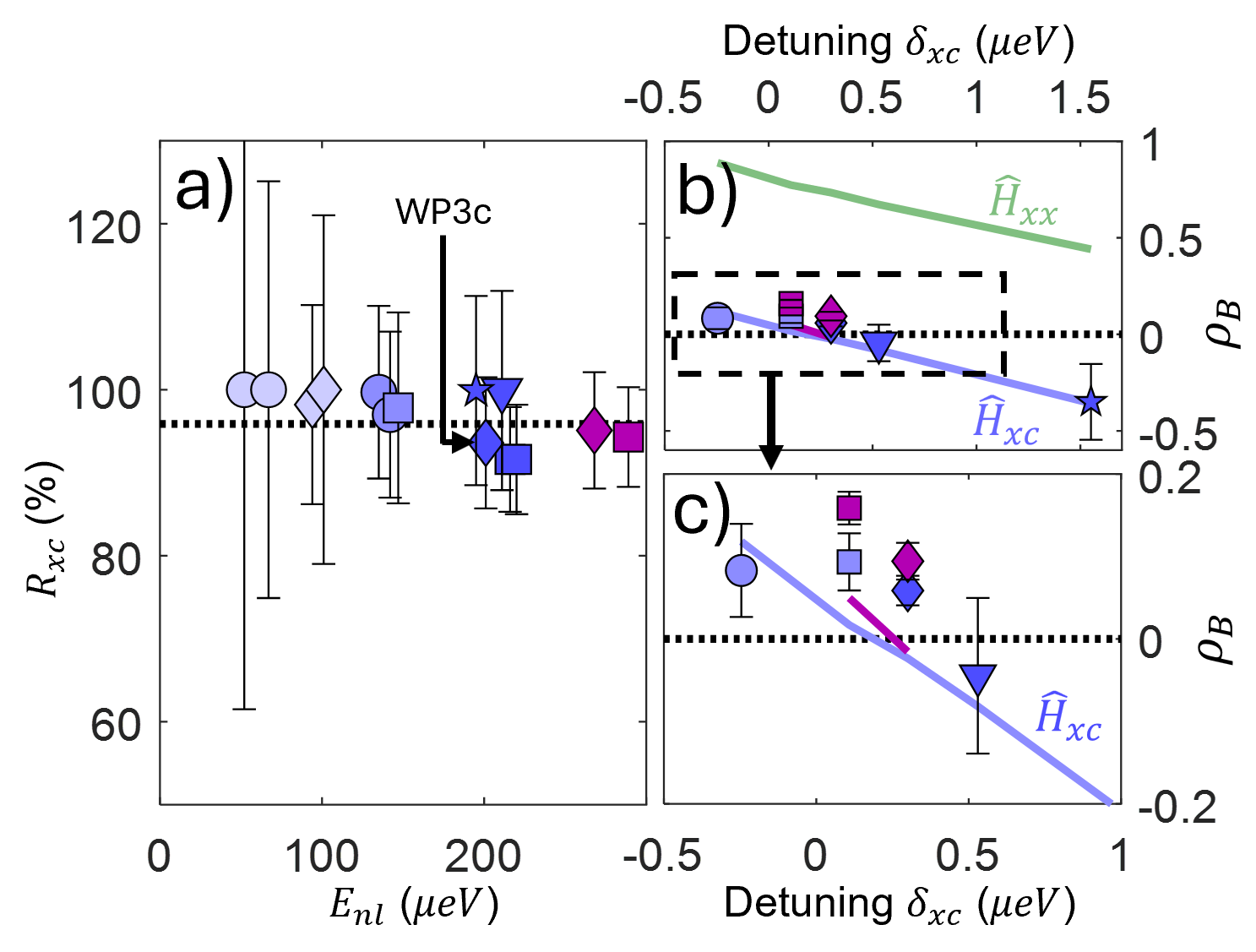}
\caption{(a) contribution of the saturation mechanism $R_{xc}=g_{xc}/(g_{xc}+g_{xx})$ to the polariton-polariton interaction as a function of the characteristic interaction energy $E_{nl}$. Five WPs with exciton-photon detunings $\delta_{xc}=-0.25$, $+0.10$, $+0.30$, $+0.53$, $+1.55$ meV are plotted as circles, squares, diamonds, triangle and star symbols respectively. For each WP, the lowest to highest probed $E_{nl}$ are plotted from light to dark shades of blue (purple for the highest). The dashed black line shows the average ratio $\langle R_{xc} \rangle=96\%$. The arrow indicates WP3c. (b) Upper- to lower-polaritons spectral shift ratio $\rho_B$ as a function of $\delta_{xc}$ using the same symbols and colors notations. A single parameter set per WP is plotted for clarity: WP1c, WP2a, WP2d, WP3c, WP3d, WP4a, WP5a (see SM Table II \cite{SuppMat}). The thick blue (green) line is the Bogoliubov theory with $R_{xc}=\langle R_{xc}\rangle$ ($R_{xc}=0\%$). (c) is a close-up of (b). WP2d and WP3d (purple color), are different from WP2a and WP3c (blue) only by their higher $E_{nl}$. The theory for WP2d and WP3d is the solid purple line segment.}
\label{fig2}
\end{figure}

% Free parameters in M_B: gn, no need for absolute knowledge of densities
The comparison of these measured dispersion relations with the predictions of the Bogoliubov theory is realized as follows. As we have already determined the linear regime parameters for each WPs and we know the value of the applied laser detuning $\delta_{\rm las}$, the only free parameters of the model are the two nonlinear energies $E_{nl,xx}=g_{xx}n_x$ and $E_{nl,xc}=g_{xc}n_x$ that the fitting algorithm is tasked to determine. It consists in a least-square optimization algorithm that is simultaneously applied to all points along the three measured dispersion relation branches. In doing so, our main target is to determine the relative contribution of $g_{xc}$ to the polariton-polariton interaction, defined as $R_{xc}=g_{xc}/(g_{xx}+g_{xc})=E_{nl,xc}/(E_{nl,xx}+E_{nl,xc})$. importantly, we do not need to determine the absolute value of $g_{xx}$ or $g_{xc}$ independently, which would require the difficult task of determining the excitonic density $n_x$ in absolute terms.  

% perform and analyze the fits
The theoretical dispersion relations obtained in this way are shown in Fig.\ref{fig1}.(e,f) for WP3c as the solid blue line. We quantify the goodness of fit using the root mean squared error $S_{\rm fit}$ where  $S_{\rm fit}^2=\langle(\omega_{j,exp}-\omega_{j,fit})^2 \rangle_j$, where $\langle.\rangle_j$ is an average running over the experimental points $\theta_j$. For WP3c, $S_{\rm fit}=27.4 \mu$eV, and the obtained parameters are: $E_{nl,xc}=188\pm 12\,\mu$eV and $E_{nl,xx}=13 \pm 17\,\mu$eV, which leads to $R_{xc}=94\% \pm 8\%$, where the uncertainty is the $1\sigma$ confidence interval. Fitting with $H_{xc}$ only, and hence with a single free parameter $g_{xc}n_x$ yields an excellent agreement as well, of $S_{\rm fit}=28 \mu$eV (see Table V in the SM \cite{SuppMat}. The saturation mechanism is thus strongly dominant over the exciton-exciton interaction. For comparison, we also fitted WP3c assuming $g_{xc}=0$, i.e. keeping only the exciton-exciton $\hat {\cal H}_{xx}$ interaction term. The result is shown as the black dashed line in Fig.\ref{fig1}.(e,f) and yields an agreement of $S_{\rm fit}=140\,\mu$eV, i.e. $\times 5$ poorer. We performed this analysis for all investigated WPs in which $\delta_{xc}=\{-0.25;0.10;0.30;0.53;1.55\}\,$meV and different parameter sets in terms of laser detuning $\delta_{\rm las}$ and the characteristic interaction energy $E_{\rm nl}\in [50,300]\,\mu$eV (see SM Table III for the results). The resulting values of $R_{xc}$ are summarized in Fig.\ref{fig2}.a: these results systematically show a large domination of the saturation mechanism (\ref{H_xc}) over the exciton-exciton interaction (\ref{H_xx}). Its average contribution thus amounts to $\langle R_{xc} \rangle=96\%\pm11\%$.

%If we include the uncertainties of all the model parameters, we obtain an even more conservative estimate of the nonlinearities: $E_{nl,xc}=184 \pm 20\,\mu$eV, $E_{nl,xx}=20\pm 30\,\mu$eV, and $R_{xc}=90 \% \pm 13\%$. 

% We confirm the conclusion by looking specifically at the rigid shifts of the UPN versus LPN
We confirm this conclusion using a second method that does not require any fitting operation, and which is focused on the particularly high sensitivity of the upper polariton Bogoliubov branch energy shift to the different interaction contributions. This shift is determined from the data as $\Delta\omega_{u}=\omega_{B,un}(0)-\omega_{0,up}(0)$, and compared with the lower polariton shift. For a meaningful comparison the latter needs to be determined in a Bogoliubov frequency range higher than $E_{nl}/\hbar$, as discussed earlier, so that the spectral shift -- like for $\delta\omega_u$ -- is not influenced by the particle-hole contributions. This condition is met for the lower polariton for $\theta_{rs}=9^\circ$ such that we define $\Delta\omega_l=\omega_{B,ln}(\theta_{rs})-\omega_{0,lp}(\theta_{rs})$. The ratio $\rho_B=\Delta\omega_u/\Delta\omega_l$ is plotted in Fig.\ref{fig2}.b (blue symbols) as a function of the exciton-photon detuning $\delta_{xc}$ and compared to the Bogoliubov theory assuming $R_{xc}=\langle R_{xc} \rangle$ for all points (solid blue line), and $R_{xc}=0$ (solid green line). The agreement with a saturation mechanism dominated interaction is again striking: in particular the sign change of $\rho_B$ around $\delta_{xc}\simeq 0$ is a strong indicator of the compensation mechanism between $\tilde(\Omega)$ and $\tilde{\omega}_x$ that occurs for the upper polariton branch when a saturation mechanism dominates as discussed earlier. Within our experimental resolution, we checked that the dependence of $\rho_B$ on $E_{nl}$ is also qualitatively consistent with our mean field model dominated by a saturation mechanism (cf. Fig.\ref{fig2}.c).

% check reservoir as alternate explanation
We conclude our study by analyzing alternative explanations of our observations. Recent works~\cite{Walker_2017,Stepanov_2019,Pieczarka_2021} have shown that, due to imperfections in the quantum well, a steady-state incoherent reservoir of dark excitons can sometimes build up and coexist with polaritons. The polariton-reservoir interaction via a scattering-like mechanism also leads to modifications of the Bogoliubov dispersion relations, and we thus examined this possibility by means of a suitable extension of our Bogoliubov model~\cite{Stepanov_2019} (see SM section II.C \cite{SuppMat}). These modifications are distinct from those resulting from the interactions discussed before, so that by fitting our experimental data, we can unambiguously determine the new free parameter $E_{nl,r}=g_rn_r$ independently from $g_{xc}n_x$ and $g_{xx}n_x$. This analysis yields for WP3c to $g_{xc}n_x=188\pm10\,\mu$eV, $g_{xx}n_x=12\pm16\,\mu$eV, and $g_rn_r=2\pm26\,\mu$eV and an unchanged  goodness of fit $S_{\rm fit}=27.4$. Performed over all WPs (see See section II.C in the SM for details \cite{SuppMat}), this analysis shows that the addition of a reservoir does not improve the goodness of fit, and that the reservoir remains always small, i.e. $g_rn_r\ll E_{nl}$. Moreover, by setting $H_{xx}$ to zero in this model (see SM table V \cite{SuppMat}) and comparing with the previous model in terms of goodness of fit, we find that both the reservoir or $H_{xx}$ can possibly account for the remaining $~5\%$ contribution to the polariton-polariton.

% more exotic interactions
A further process was proposed in \cite{Grudinina_2024} where the polariton reservoir could interact with polaritons also by saturating the excitonic oscillator strength. This hypothesis either yields a poor fit with the data or, when combined with exciton-exciton scattering, the fit converges on unrealistic Bogoliubov states (see section II.D.1 and II.E in the SM \cite{SuppMat}). Finally, we considered two other possible interaction mechanisms: exciton-photon scattering, and a nonlinear contribution to the Rabi coupling. A quantitative comparison with the data (see section II.D.2-3 and II.E in the SM \cite{SuppMat})) shows that neither of them offers a better fit to the data than $H_{xc}$. 

Let us highlight that our conclusion is fully compatible with previous works where only the exciton-exciton interaction term $\hat {\cal H}_{xx}$ was used to model polariton-polariton interactions: these works focused on the lower polariton dispersion branch only, in which both terms $\hat {\cal H}_{xx}$ and $\hat {\cal H}_{xc}$ lead to repulsive $lp$-$lp$ interactions, which is the key feature in polaritonic quantum fluids phenomena. Only this dedicated analysis -- that crucially involves the upper polariton branch -- has allowed us to clearly distinguish between these two mechanisms. The consequences of this distinction shows up at a more fundamental level: for instance the dominance of $H_{xc}$ suggests that the polariton-polariton interaction strength can be enhanced by increasing the light-matter coupling.

Note also that in this work we have not made any assumptions on the microscopic mechanisms underlying the $H_{xx}$ and $H_{xc}$ Hamiltonian terms. To shed light on them, different microscopic theories are available in the literature, e.g. accounting for the electron-hole degrees of freedom~\cite{Combescot_2007,Combescot_2008b} and their coupling to photons~\cite{Bleu_2020,Li_2021,Nakano_2024,Grudinina_2024}. Interestingly, in \cite{Li_2021,Christensen_2022}, the authors observe in their model some features in the lower polariton sector that are consistent with a saturation mechanism; however, as discussed earlier, a meaningful comparison between experiment and theory needs addressing both the lower and upper polaritons sectors. In a forthcoming work, we will extend the model developed in \cite{Christensen_2022} to explicitly include the upper polariton branch as well as saturation effects, and compare the predictions to the experimental observations~\cite{in_preparation}.
 
%conclusion
In conclusion, our results demonstrate that, in a widely-used InGaAs-based microcavity design, the polariton-polariton interaction is dominated by excitonic oscillator-strength saturation effects while exciton-exciton scattering or the effect of an incoherent reservoir  contribute to less than $5\%$ of the total interaction energy over a wide range of exciton-photon detuning and polariton density. This result establishes a saturation-based picture of polariton-polariton interactions with far reaching consequences in the quest for reaching the full quantum regime.

{\it Acknowledgment} -- MR, FC and YZ acknowledge financial support from the Centre for Quantum technologies 'Exploratory Initiative program', and from the National Research Foundation via CNRS@CREATE internal grant 'NGAP'. IC acknowledges financial support from the Provincia Autonoma di Trento, from the Q@TN Initiative, and from the National Quantum Science and Technology Institute through the PNRR MUR Project under Grant PE0000023-NQSTI, co-funded by the European Union - NextGeneration EU.

\pagebreak

\onecolumngrid
\begin{center}
  \textbf{\large SUPPLEMENTARY INFORMATION \\ Excitonic oscillator-strength saturation dominates polariton-polariton interactions}\\[1cm]
\end{center}

\section{Non-interacting polaritons: Non-resonant-photoluminescence}

\begin{figure}[h!]
\includegraphics[width=0.5\columnwidth]{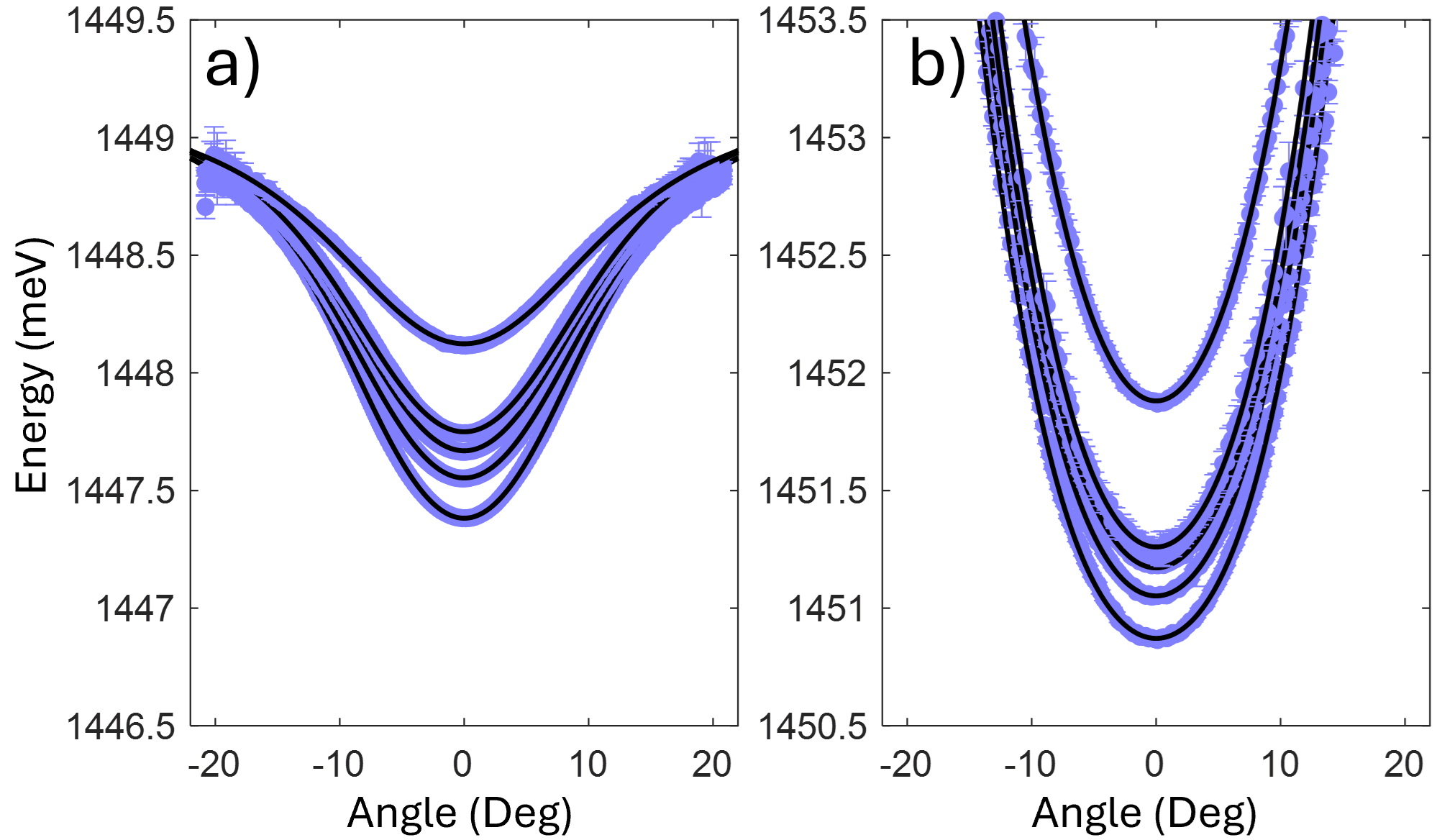}
\caption{Measured (round symbols) and fitted (black line) lower- (a) and upper- (b) polaritons dispersion relation. In both panels the dispersion relations of WP1 to WP5 are given by the curves from lower to higher energy (at $\theta=0$).}
\label{fig_SM1}
\end{figure}

The dispersion relations of the non-interacting polaritons are obtained by exciting a weak density of free electron holes $55\,$meV above the excitonic transition, that relax efficiently into the polariton states. We fit the measurements using a coupled harmonic oscillators model, that correspond to the bracket terms in $\hat{\cal H}_0$ [cf. Eq.\ref{eq:H0}]. The comparison between the measured and fitted dispersion relations are shown in Fig.\ref{fig_SM1}, and the corresponding results for each working points (WPs) investigated in this work are summarized in Table \ref{tab:linear_parameters}.

\begin{table}[h!]
    \centering
    \begin{tabular}{|c||c|c|c|c|}
           \hline
         WP  & $\hbar\Omega$ [meV] & $\hbar\omega_x$ [meV] & $\delta_{xc}$ [meV] & $n_{cav}$\\
           \hline
      WP1  & $3.48\pm 0.02$ & $1449.25\pm 0.01$ &  $-0.25 \pm 0.02$ & $3.40 \pm 0.04$    \\
      WP2  & $3.50\pm 0.02$ & $1449.25\pm 0.01$ &  $ 0.11 \pm 0.02$ & $3.43 \pm 0.04$    \\
      WP3  & $3.49 \pm 0.02$ & $1449.27\pm 0.01$ &  $ 0.30 \pm 0.03$ & $3.42 \pm 0.05$    \\
      WP4  & $3.47 \pm 0.03$ & $1449.24\pm 0.01$ &  $ 0.53 \pm 0.02$ & $3.38 \pm 0.03$    \\
      WP5  & $3.42 \pm 0.02$ & $1449.22\pm 0.01$ &  $ 1.55 \pm 0.02$ & $3.46 \pm 0.03$    \\
      \hline
    \end{tabular}
    \caption{Working points parameters obtained from the photoluminescence (non-interacting polaritons) analysis.}
    \label{tab:linear_parameters}
\end{table}

\section{Interacting polaritons}
\subsection{Details of the model}
A similar version of this derivation can be found in \cite{Frerot_2023}. 
We recall that the expression for the non-interacting Hamiltonian is: 
\begin{equation}
\hat{\cal H}_0 =  \hbar \sum_{\bf k} \left[ \omega_x \hat b_{\bf k}^\dagger \hat b_{\bf k} + \omega_{c,{\bf k}} \hat a_{{\bf k}}^\dagger \hat a_{{\bf k}} + \frac{\Omega}{2} (\hat a_{\bf k}^\dagger \hat b_{\bf k} + {\rm h.c.})\right] + ({\hat a}_{\bf k_p}F_0e^{i\omega_{\rm las}t} + {\rm h.c.})
\label{eq:H0}
\end{equation}
We obtain the equations of motion (EoM) for the field using $i \hbar \partial_t \hat O = [\hat O, \hat{\cal H}]$ for $\hat O=\hat a_{\bf k}, \hat b_{\bf k}$, and $\hat{\cal H}= \hat{\cal H}_0 + \hat{\cal H}_{int}$, with $\hat{\cal H}_{int}$ describing various possibilities for the interaction among polaritons. Without interactions ($\hat{\cal H}_{int}=0$), the EoM read:
\begin{eqnarray}
    i\partial_t \hat a_{\bf k} &=& \omega_{c,{\bf k}} \hat a_{\bf k} + \frac{\Omega}{2} \hat b_{\bf k} +  \delta_{{\bf k}, {\bf k_p}} F_0 e^{-i\omega_{\rm las}t} \\
    i\partial_t \hat b_{\bf k} &=& \omega_{x} \hat b_{\bf k} + \frac{\Omega}{2} \hat a_{\bf k}
\end{eqnarray}
where the pump term $F_0$ is only coupled to the photon field $\hat a_{\bf k_p}$ at the pump wavevector ${\bf k}={\bf k_p}$. 
We then move to the frame rotating at the drive frequency, defining the operators $\hat O^{({\rm rot})} = \hat O e^{i\omega_{\rm las}t}$. The rotating operators obey the EoM $i\partial_t \hat O^{({\rm rot})} = e^{i\omega_{\rm las}t} i\partial_t \hat O - \omega_{\rm las} \hat O^{({\rm rot})}$. From now on, we work in the rotating frame, but omit the $({\rm rot})$ superscript for simplicity of notation. Hence the non-interacting EoM:
\begin{eqnarray}
    i\partial_t \hat a_{\bf k} &=& (\omega_{c,k} - \omega_{\rm las}) \hat a_{\bf k} + \frac{\Omega}{2} \hat b_{\bf k} + F_0 \delta_{{\bf k}, {\bf k_p}}\\
    i\partial_t \hat b_{\bf k} &=& (\omega_{x} - \omega_{\rm las}) \hat b_{\bf k} + \frac{\Omega}{2} \hat a_{\bf k}
\end{eqnarray}

We then add to the model various interaction terms, starting from $\hat H_{xx}$ and $\hat H_{xc}$ on which we focus in the main text. We then discuss the possible inclusion of a reservoir of dark-excitons in Section \ref{sec_reservoir}. We finally discuss the inclusion of further possible terms $\hat H_3$ and $\hat H_4$ in Sections \ref{sec_H3} and \ref{sec_H4} respectively. All interaction terms are fourth-order in the polariton operators, and we recall the expressions for $\hat H_{xx}$ as given in the main text:
\begin{equation}
\hat {\cal H}_{xx} = \frac{\hbar g_{xx}}{2} \sum_{{\bf k},{\bf k}',{\bf q}}
\hat b_{{\bf k}+{\bf q}}^\dagger \hat b_{{\bf k}'-{\bf q}}^\dagger \hat b_{{\bf k}'} \hat b_{\bf k}
\label{H_xx}
\end{equation}
and for $\hat H_{xc}$:
\begin{equation}
\hat {\cal H}_{xc} = -\frac{\hbar g_{xc}}{2}\sum_{{\bf k},{\bf k}',{\bf q}} (\hat a_{{\bf k}+{\bf q}}^\dagger \hat b_{{\bf k}'-{\bf q}}^\dagger \hat b_{{\bf k}'} \hat b_{\bf k} + {\rm h.c.}),
\label{H_xc}
\end{equation}
The first step of the analysis is to derive the mean-field solution, namely focusing only on the fields at the drive wavevector ${\bf k}={\bf k}_p$, and making the mean-field approximation $\hat a_{\bf k_p} \approx \langle \hat a_{\bf k_p} \rangle := \psi_c$ and $\hat b_{\bf k_p} \approx \langle \hat b_{\bf k_p} \rangle := \psi_x$.
The physical assumption is then that the occupation of all other modes is much less than the mean-field occupation.
The second step is then to focus on all other wavevectors ${\bf k} \neq {\bf k_p}$, and write an effective quadratic Hamiltonian by keeping only the terms up to second order on the corresponding operators. 

Formally, the interaction Hamiltonian involves terms of the form:
\begin{equation}
    \hat V = V_0\sum_{{\bf k}, {\bf k'}, {\bf q}} \hat w_{\bf k + q}^\dagger \hat x_{\bf k'-q}^\dagger \hat y_{\bf k'} \hat z_{\bf k} 
\end{equation}
with $\hat w_{\bf k}, \hat x_{\bf k}, \hat y_{\bf k}, \hat z_{\bf k}$ replaced by either photon operators $\hat a_{\bf k}$ or exciton operators $\hat b_{\bf k}$. The corresponding effective quadratic term is obtained by keeping only terms where two labels are equal to the pump wavevector ${\bf k_p}$, and replacing the corresponding operators by their mean-field value, that we denote as $\langle \hat w_{\bf k_p} \rangle = \psi_w$, $\langle \hat x_{\bf k_p} \rangle = \psi_x$, $\langle \hat y_{\bf k_p} \rangle = \psi_y$ and $\langle \hat z_{\bf k_p} \rangle = \psi_z$. We then obtain six quadratic terms:
\begin{equation}
    \hat V^{(2)} = V_0\sum_{\bf q} \left( \hat w_{\bf k_p+q}^\dagger \hat x_{\bf k_p-q}^\dagger \psi_y \psi_z + \hat w^\dagger_{\bf q} \hat y_{\bf q} \psi_x^* \psi_z + \hat x_{\bf q}^\dagger \hat y_{\bf q} \psi^*_w \psi_z + \hat w_{\bf q}^\dagger \hat z_{\bf q} \psi_x^* \psi_y + \hat x_{\bf q}^\dagger \hat z_{\bf q} \psi_w^* \psi_y + \hat y_{\bf k_p+q} \hat z_{\bf k_p-q} \psi_w^* \psi_x^* \right)
\end{equation}
These various terms may be interpreted as giving rise to effective contributions to the Rabi splitting $\hbar \Omega$, or to the bare energy of the modes.

\subsection{Bogoliubov excitations assuming interactions described by $\hat H_{xx}$ and $\hat H_{xc}$}
The mean-field EoM with the interaction terms $\hat H_{xx}$ and $\hat H_{xc}$ read:
\begin{equation}
    i\frac{\partial \psi_x}{\partial t}=\left(\omega_x-\omega_{\rm las}-i\frac{\gamma_x}{2}+g_{xx}|\psi_x|^2\right)\psi_x+\left(\frac{\Omega}{2}-g_{xc}|\psi_x|^2\right)\psi_c-\frac{g_{xc}}{2}\psi_x^2\psi_c^*  
    \label{MFx}
\end{equation}
and:
\begin{equation}
    i\frac{\partial \psi_c}{\partial t}=\left(\omega_{c,\mathbf{k_p}}-\omega_{\rm las}-i\frac{\gamma_{c}}{2}\right)\psi_c+\left(\frac{\Omega}{2}-\frac{g_{xc}}{2}|\psi_x|^2\right)\psi_x+F_{0}~,
    \label{MFc}
\end{equation}
with $\gamma_x$ and $\gamma_c$ phenomenological damping rates. 
Owing to our continuous wave (CW) laser drive, we can take the steady-steady limit of these equations, and thus obtain an analytic expression for the excitonic and photonic mean fields $\psi_x=\sqrt{n_x}$ and $\psi_c=\sqrt{n_c}e^{-i\phi_{xc}}$ as a function of the system parameters, where we have dropped a global phase, and $\phi_{xc}$ is the relative phase between the two fields. We can then determine the Bogoliubov excitation around this steady-state by expanding $\hat{\cal H}$ to quadratic order in the excitonic and photonic operators of momenta $\mathbf{k}\neq\mathbf{k_p}$. Their equations of motions in the Heisenberg picture can be expressed as
\begin{equation}
i\partial \hat{A}/\partial t=\hat{M}_B \hat{A}
\end{equation}
where $\hat{A}=[\hat a_{\bf k}, \hat b_{\bf k}, \hat a_{\bf 2k_p-k}^\dagger, \hat b_{\bf 2k_p-k}^\dagger]^T$ is the excitonic and photonic fluctuations vector, and $\hat{M}_B({\bf k_p},{\bf k})$ is The $4{\times} 4$ Bogoliubov matrix. Its eigenvalues provide the 4 dispersion relations branches $\{\omega_{B,ug}({\bf k}),\omega_{B,lg}({\bf k}),\omega_{B,ln}({\bf k}),\omega_{B,un}({\bf k})\}$ of the Bogoliubov excitations. The explicit expression of the Bogoliubov matrix is
\begin{equation}
    \hat{M}_B = \begin{pmatrix}
        \hat{m}_{{\rm P},\mathbf{k}}   &   \hat{m}_{\rm HP}  \\ 
        \hat{m}_{\rm PH}  &   \hat{m}_{{\rm H},\mathbf{k},\mathbf{k_p}}
    \end{pmatrix} ~,
    \label{eq_MBstructure}
\end{equation}
where 
\begin{equation}
    \hat{m}_{{\rm P}, {\bf k}} = \begin{pmatrix}
        \omega_{c,\mathbf{k}}-\omega_{\rm las}-i\gamma_c/2 & \Omega/2 - g_{xc}n_x  \\ 
        \Omega/2 - g_{xc}n_x              & \omega_{x} -\omega_{\rm las}-i\gamma_x/2 + 2g_{xx}n_x - 2g_{xc}\sqrt{n_xn_c}\cos{\phi_{xc}}
    \end{pmatrix} ~,
    \label{eq_mP_xx_xc}
\end{equation}
and
\begin{equation}
    \hat{m}_{\rm HP} = \begin{pmatrix}
        0            & -g_{xc}n_x/2  \\ 
        -g_{xc}n_x/2 & g_{xx}n_x - g_{xc}\sqrt{n_xn_c}e^{-i\phi_{xc}} 
    \end{pmatrix} ~,
    \label{eq_MB}
\end{equation}
and $\hat{m}_{\rm PH}=-\hat{m}_{\rm HP}^*$, and $\hat{m}_{{\rm H},\mathbf{k}}=-\hat{m}_{{\rm P},2\mathbf{k_p}-\mathbf{k}}^*$. Notice that the actual frequencies are obtained by shifting them back to the laboratory frame, namely adding $\omega_{\rm las}$ to all of them. 

\subsubsection{Interaction strengths determined from the comparison between the experiments and the model}

\begin{figure}[h]
\includegraphics[width=0.9\textwidth]{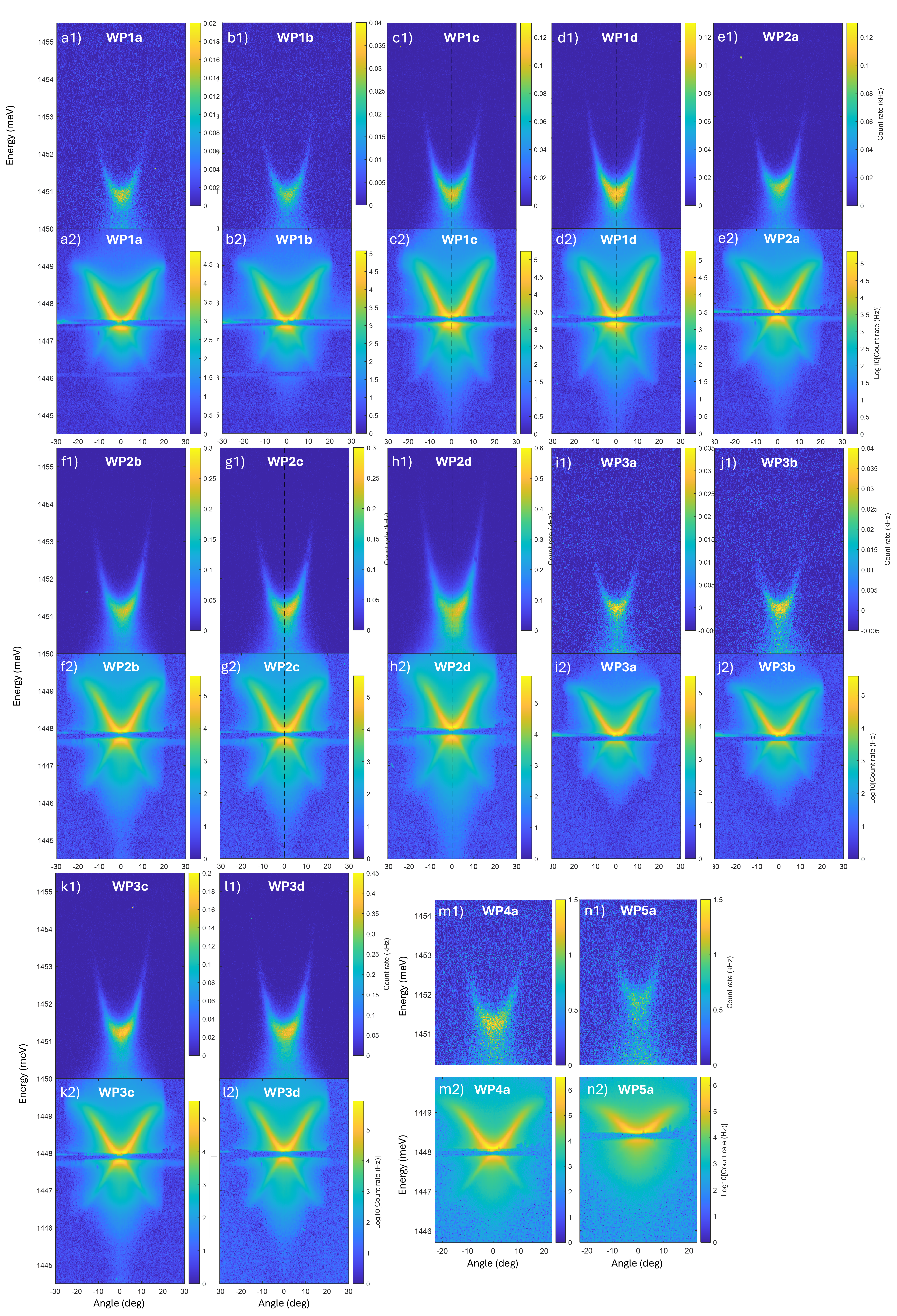}
\caption{Raw measurements of $I(\omega,\theta)$ for all investigated parameters WP$nj$ as listed in Table (\ref{tab:NL_param}).}
\label{fig_SM2}
\end{figure}

\begin{figure}[h]
\includegraphics[width=0.95\textwidth]{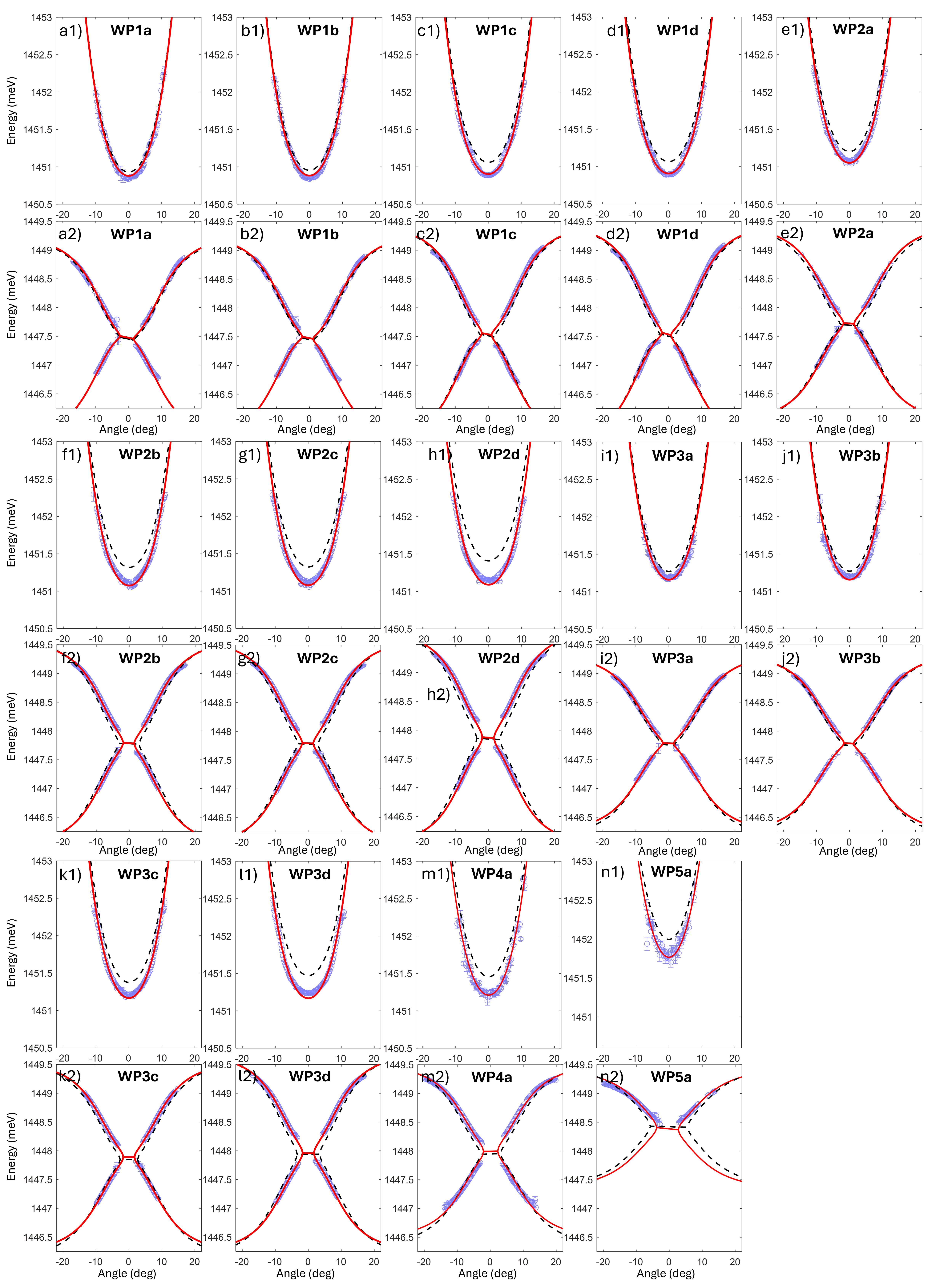}
\caption{labels Measured (blue circles) and fitted (solid and dashed lines) Bogoliubov dispersion relation for for all investigated parameters WP$nj$ as listed in Table (\ref{tab:NL_param}). The solid red line is the Model that includes both $H_{xx}$ and $H_{xc}$ (corresponding numerical results in Table \ref{tab:coulomb_sat}). The dashed black line is the theory assuming $H_{xx}$ only.}
\label{fig_SM3}
\end{figure}

The measured raw emission maps $I_B(\theta,\omega)$ of the Bogoliubov excitations are shown in Fig.\ref{fig_SM2} for all the investigated experimental parameters. From them are extracted the measured Bogoliubov dispersion relations shown in Fig.\ref{fig_SM3} (blue circle). The fitted theory is shown as red lines in Fig.\ref{fig_SM3}, and the corresponding fitted parameters are listed in Table (\ref{tab:coulomb_sat}). The two interactions strength determined from the fit are $g_{xx}n_x$ and $g_{cx}n_x$, and $R_{xc}$ is the relative contribution of $g_{cx}n_x$ to the total interaction strength.

Note that in WP1 (panels a2, b2, c2 and d2 in Fig.\ref{fig_SM3}), in spite of the fact that we excite a polaritonic mean field with zero in-plane momentum, we observe Bogoliubov dispersion relations \textit{lg} and \textit{ln} that are characteristics of a small non-zero in-plane momentum (manifested by a broken up-down mirror symmetry between the normal and the ghost branches). This is due to the built-in in-plane gradient $\nabla\omega_{c,0}|$ of the bare cavity resonance. That turns into a polaritonic  energy gradient $|\nabla\omega_{0,lp}|=|C|^2|\nabla\omega_{c,0}|$, where $|C|^2$ is the photonic fraction. WP1 is the most negatively-detuned working point ($\delta_{xc}=-0.25\,$meV, $|C|^2=54\%$) investigated in this work (cf. table (\ref{tab:linear_parameters})), and the corresponding energy gradient experienced by polaritons within the excitation spot results in a small in-plane momentum along this gradient. This is taken into account in the theoretical fits of WP1 by including this small momentum into the pump laser (the largest such corrections amounts to $\theta_p=0.66^\circ$, where $\theta_p=0.66^\circ$ is the  effective incidence angle of the laser [where nominally, $\theta_p=0^\circ$]. ). For the other WPs, the energy gradient across the spot is essentially negligible.

\begin{table}[h!]
    \centering
    \begin{tabular}{|c||c|c|}
          \hline
        WP${jk}$ Label  & $\delta_{\rm las}$ [$\mu$eV] & $P_{las}$ [mW] \\ 
          \hline
      WP1a & $101\pm 10$ & 2.8 \\
      WP1b & $101\pm 10$ & 4.8 \\
      WP1c & $160\pm 10$ & 19.0 \\
      WP1d & $160\pm 10$ & 16.6 \\
      WP2a & $167\pm 10$ & 9.0  \\
      WP2b & $251\pm 10$ & 15.5 \\
      WP2c & $251\pm 10$ & 17.5 \\
      WP2d & $319\pm 10$ & 35.0 \\
      WP3a & $110\pm 10$ & 4.0 \\
      WP3b & $110\pm 10$ & 5.1 \\
      WP3c & $203\pm 10$ & 13.7 \\
      WP3d & $296\pm 10$ & 32 \\
      WP4a & $228\pm 10$ & - \\
      WP5a & $293\pm 10$ & - \\ 
      \hline
    \end{tabular}
    \caption{Experimental conditions for every measurements of Bogoliubov excitation dispersion relations. Each experiment WP$_{jk}$ is labeled by its working point "WP$_j$" and a different letter $k$. Note that the laser power $P_{las}$ is irrelevant in the comparison between theory and experiment, i.e. it is not used or needed in the fitting procedure. '-' indicated missing data.}
    \label{tab:NL_param}
\end{table}

\begin{table}[h!]
    \centering
    \begin{tabular}{|c||c|c|c|c|}
          \hline
        WP${jk}$ Label & $g_{cx}n_x$ [$\mu$eV] & $g_{xx}n_x$ [$\mu$eV] & $R_{xc}$ [$\%$] & $S_{\rm fit}\,$[$\mu$eV] \\ 
          \hline
      WP1a & $52\pm 14$ &  $0 \pm 20$ & $100 \pm 36$ & 33.6 \\
      WP1b & $67\pm 11$ &  $0 \pm 17$ & $100 \pm 25$ & 28.4 \\
      WP1c & $135\pm 9$ &  $0 \pm 14$ & $100 \pm 10$ & 23.9 \\
      WP1d & $138\pm 10$&  $4 \pm 15$ & $97 \pm 10$ & 25.9 \\
      WP2a & $144\pm 12$ &  $3 \pm 17$ & $98 \pm 11$ & 24.7 \\
      WP2b & $198\pm 10$ &  $18 \pm 15$ & $92 \pm 6$ & 26.6 \\
      WP2c & $202\pm 11$ &  $18 \pm 16$ & $92 \pm 7$ & 28.4 \\
      WP2d & $273\pm 12$ &  $16 \pm 18$ & $94 \pm 6$ & 34 \\
      WP3a & $93\pm 8  $ &  $2 \pm 12$  & $98\pm 12$ & 22.8 \\
      WP3b & $101\pm 15$ &  $0 \pm 21$  & $100\pm 21$ & 37.8 \\
      WP3c & $188\pm 12$ &  $13 \pm 17$  & $94\pm 8$ & 27.4 \\
      WP3d & $255\pm 14$ &  $13 \pm 20$  & $95\pm 7$ & 35.9 \\
      WP4a & $212\pm 20$ & $0\pm 26$ & $100\pm 12$ & 58.4 \\
      WP5a & $221\pm 26$ & $15\pm 27$ & $94\pm 11$ & 70.1 \\ 
      \hline
    \end{tabular}
    \caption{Results of the comparison between the measured Bogoliubov dispersion relation and the $H_{xx}$ and $H_{xc}$ model. WP3c is the experiment presented in detail in the main text. Missing data are marked as '-'. Note that the laser power $P_{las}$ is irrelevant in the comparison between theory and experiment, i.e. it is not used or needed in the fitting procedure.0}
    \label{tab:coulomb_sat}
\end{table}

\begin{table}[h!]
    \centering
    \begin{tabular}{|c||c|c|c|c|c|}
    \hline
    WP${jk}$ label & $g_{cx}n_x$ [$\mu$eV] & $g_{xx}n_x$ [$\mu$eV] & $g_rn_r$ [$\mu$eV] & $S_{\rm fit}\,$[$\mu$eV] \\ \hline
      WP1a  & $52\pm 14$  & $0\pm19$ &  $0\pm35$ & 33.6\\
      WP1b  & $67\pm 4$   & $0\pm7$  &  $0\pm14$ &  28.4 \\
      WP1c  & $134\pm 8$  & $0\pm13$ &  $0\pm22$  & 23.9 \\
      WP1d  & $138\pm 9$  & $1\pm14$ &  $6\pm23$  & 25.8 \\   
      WP2a  & $144\pm 12$ & $1\pm16$ &  $5\pm26$  & 24.6 \\
      WP2b  & $192\pm 10$ & $0\pm15$ &  $39\pm21$  & 25.9 \\
      WP2c  & $197\pm 11$ & $0\pm15$ &  $38\pm22$  & 28 \\
      WP2d  & $268\pm 12$ & $0\pm18$ &  $37\pm25$  & 33.2 \\
      WP3a  & $91\pm 8$   & $0\pm12$ &  $7\pm17$  & 22.8 \\ 
      WP3b  & $101\pm 15$ & $0\pm21$ &  $0\pm37$  & 37.8 \\ 
      WP3c & $188\pm 12$ & $12\pm16$&  $2\pm26$  & 27.4 \\ 
      WP3d  & $250\pm 10$ & $0\pm13$&   $28\pm19$  & 35.6 \\ 
      WP4a  & $212\pm19$  & $0\pm25$ &  $0\pm38$  & 58.4 \\
      WP5a  & $196\pm20$  & $0\pm22$ &  $54\pm24$  & 67.5 \\
      \hline
    \end{tabular}
    \caption{Results of the comparison between the measured Bogoliubov dispersion relation and the model that include $H_{xx}$,$H_{xc}$, and the reservoir (scattering interaction version).}
    \label{tab:coulomb_sat_res}
\end{table}

\begin{table}[h!]
    \centering
    \begin{tabular}{|c||c|c|c|c|c|}
    \hline
    WP${jk}$ label & $g_{xc}n_x$ [$\mu$eV] & $g_rn_r$ [$\mu$eV] & $S_{\rm fit}\,$[$\mu$eV] \\ \hline
      WP1a  & $52\pm 14$  & $ 0\pm 38$ &  $33.6$ \\
      WP1b  & $67\pm 11$  & $ 0 \pm 31 $ & $28.4$ \\
      WP1c  & $135\pm 9$  & $1\pm 24$ &  $23.8$ \\
      WP1d  & $138\pm 9$  & $8\pm 25$ &  $25.8$ \\   
      WP2a  & $144\pm 11$  & $2\pm 27$ &  $24.6$ \\
      WP2b  & $193\pm 10$  & $35\pm 21$ &  $25.9$ \\
      WP2c  & $197\pm 11$  & $34 \pm 23$ &  $28$ \\
      WP2d  & $268\pm 12$  & $33\pm 25$ &  $33.2$ \\
      WP3a  & $93\pm 7$  & $3\pm 16$ &  $22.8$ \\ 
      WP3b  & $101\pm 15$  & $0\pm 35$ &  $37.8$ \\
      WP3c  & $188\pm 10$  & $21\pm 24$ &  $27.5$ \\ 
      WP3d  & $251\pm 9$  & $28\pm 23$ &  $35.5$ \\
      WP4a  & $212\pm 20$  & $0\pm 40$ &  $58.4$ \\
      WP5a  & $196\pm 35$  & $54\pm 50$ &  $66.9$ \\
      \hline
    \end{tabular}
    \caption{Results of the comparison between the measured Bogoliubov dispersion relation and the model that includes $H_{xc}$ and the reservoir (scattering interaction version).}
    \label{tab:sat_res}
\end{table}

\newpage
\subsection{Bogoliubov excitations assuming interactions described by $\hat H_{xx}$, $\hat H_{xc}$a and polariton scattering with an incoherent reservoir}
\label{sec_reservoir}

Due to imperfections in the quantum wells, an incoherent reservoir of dark excitons can sometimes build up, feeding off, and co-existing with polaritons. This reservoir can be added to our model as described in \cite{Stepanov_2019}. The resulting MF equation for the photonic field is identical to Eq.~(\ref{MFc}), while that of the excitonic field involves an additional spectral shift caused by the reservoir and becomes
\begin{equation}
    i\frac{\partial \psi_x}{\partial t}=\left(\omega_x-\omega_{\rm las}-i\frac{\gamma_x}{2}+g_{xx}|\psi_x|^2+g_rn_r\right)\psi_x+\left(\frac{\Omega}{2}-g_{xc}|\psi_x|^2\right)\psi_c-\frac{g_{xc}}{2}\psi_x^2\psi_c^*~, 
    \label{MFx_with_reservoir}
\end{equation}
where $\hbar g_r$ is the exciton-reservoir interaction strength and $n_r$ is the reservoir density. The reservoir density is described by a third equation: 
\begin{equation}
    \frac{\partial n_r}{\partial t}=-\gamma_r n_r+\gamma_x|\psi_x|^2~, 
    \label{MF_reservoir}
\end{equation}
describing an effective reservoir population dynamics, where $\gamma_r$ is the reservoir decay rate and $\gamma_x$ is the transfer rate of bright excitons (those involved in the polaritonic state) into the reservoir.  We introduce $\rho_r^2=n_r$, which obeys the equation:
\begin{equation}
    \partial_t \rho_r = \frac{1}{2\rho_r} \partial_t n_r = -\frac{\gamma_r}{2} \rho_r + \frac{\gamma_x n_x}{2 \rho_r} ~.
    \label{eq_rhor}
\end{equation}

Similarly to the previous sections, taking the steady state limit, the excitonic and photonic mean-fields can be determined as well as the reservoir steady state population $n_r$. We can then derive the equation of motions of the fluctuations as  
\begin{equation}
i\partial \hat{A_r}/\partial t=\hat{M}_{B,r} \hat{A_r}
\end{equation}
where $\hat{A_r}=[\delta\psi_{c,{\bf k}},\delta\psi_{x,{k}},\delta\psi_{c,{\bf 2 k_p - k}}^*,\delta\psi_{x,{\bf 2k_p-k}}^*,\delta\rho_r]^T$ is the fluctuations vector, where $\langle a_{\mathbf{k}\neq\bf{k_p}}\rangle=\delta\psi_{c,{\bf k}}$, $\langle b_{\mathbf{k}\neq\bf{k_p}}\rangle=\delta\psi_{x,{\bf k}}$, and $\delta\rho_r$ are the reservoir fluctuations. $\hat{M}_{B,r}$ is the $5\times 5$ Bogoliubov matrix that reads:
\begin{equation}
    \hat{M}_{B,r} = \begin{pmatrix}
        \hat{m}_{{\rm Pr},\mathbf{k}}   &   \hat{m}_{\rm HP} & \hat{m}_{r1}\\ 
        \hat{m}_{\rm PH}  &   \hat{m}_{{\rm Hr},\mathbf{k}} & -\hat{m}_{r1}\\
        \hat{m}_{r2}      &     \hat{m}_{r2}                & -i\gamma_r' \\
    \end{pmatrix} ~,
    \label{eq_MBr}
\end{equation}
where $\hat{m}_{r1}=[0,2g_r\sqrt{n_xn_r}]^T$, $\hat{m}_{r2}=[0,i\frac{\gamma_x}{2}\sqrt{n_x/n_r}]$, and $\gamma_r' = \frac{1}{2}(\gamma_r + \gamma_x n_x / n_r) = \gamma_r$ (since by Eq.~\eqref{eq_rhor}, in the steady state $\gamma_r n_r = \gamma_x n_x$). We have introduced $\hat m_{{\rm Pr}, {\bf k}} = \hat m_{{\rm P}, {\bf k}} + \begin{pmatrix}
    0 & 0 \\ 0 & g_r n_r
\end{pmatrix}$, and $\hat m_{{\rm Hr}, {\bf k}} = -\hat m_{{\rm Pr}, {\bf 2k_p-k}}^*$, with $\hat m_{{\rm P}, {\bf k}} $ given by Eq.~\eqref{eq_mP_xx_xc}. In terms of fitting the experimental data, this model involves three free parameters $g_{xx}n_x$, $g_{xc}n_x$ and $g_rn_r$, that all act upon the shape and energies of the Bogoliubov dispersion relations in a different way. As a result these parameters are well independent from each other. 

The results of the comparison between the experiment and this model are summarized in Table \ref{tab:coulomb_sat_res}. As shown by comparing the fit qualities $S_{\rm fit}$, this model does not yield any improvement with respect to the reservoir-free model (Table \ref{tab:coulomb_sat}).  Interestingly, we can also set $H_{xx}$ to zero in this model, and compare the quality of the fits assuming $H_{xc}$ and $H_{xx}$ on one hand, and $H_{xc}$ and the reservoir on the other hand. The experiment-theory comparison for the latter is summarized in table \ref{tab:sat_res}. This analysis shows that both the reservoir, or $H_{xx}$, can equally account for the remaining $\sim 5\%$ contribution which is not due to $H_{xc}$.

\subsection{Bogoliubov excitations assuming other possible interaction mechanisms}

\subsubsection{saturation-inducing reservoir hypothesis}

As suggested recently, an incoherent reservoir could also possibly act as a saturating medium for the exciton-photon interaction strength $\hbar\Omega$ \cite{Grudinina_2024}. An interaction term for this mechanism can be derived in a way similar to the previous section. The reservoir equation of motions remain the same as Eq.(\ref{MF_reservoir}), while the excitonic and photonic EOM are modified into: 

\begin{equation}
    i\frac{\partial \psi_x}{\partial t}=\left(\omega_x-\omega_{\rm las}-i\frac{\gamma_x}{2}\right)\psi_x+\left(\frac{\Omega}{2}-g_r^{sat}n_r\right)\psi_c~, 
    \label{MFx_with_rsat}
\end{equation}
and
\begin{equation}
    i\frac{\partial \psi_c}{\partial t}=\left(\omega_c-\omega_{\rm las}-i\frac{\gamma_c}{2}\right)\psi_c+\left(\frac{\Omega}{2}-g_r^{sat}n_r\right)\psi_x~, 
    \label{MFc_with_rsat}
\end{equation}

\subsubsection{$H_3$: Exciton-photon scattering}
\label{sec_H3}
The interaction Hamiltonian that we consider here is
\begin{equation}
\hat {\cal H}_{3} = (\hbar W/2) \sum_{{\bf k},{\bf k}',{\bf q}} \left(
\hat a_{{\bf k}+{\bf q}}^\dagger \hat b_{{\bf k}'-{\bf q}}^\dagger \hat a_{{\bf k}'} \hat b_{\bf k}+ {\rm h.c.} \right)
\label{H_3}
\end{equation}
Assuming only this interaction term in the model, the MF equations become: 
\begin{equation}
    i\frac{\partial \psi_x}{\partial t}=\left(\omega_x-\omega_{\rm las}-i\frac{\gamma_x}{2}+W|\psi_c|^2\right)\psi_x+\frac{\Omega}{2}\psi_c  
    \label{MFx_H3}
\end{equation}
and:
\begin{equation}
    i\frac{\partial \psi_c}{\partial t}=\left(\omega_{c,\mathbf{k_p}}-\omega_{\rm las}-i\frac{\gamma_{c}}{2}+W|\psi_x|^2\right)\psi_c+\frac{\Omega}{2}\psi_x+F_{\rm las}~.
    \label{MFc_H3}
\end{equation}
The corresponding $4\times 4$ Bogoliubov matrix $\hat{M}_{B,H3}$ has the same structure as $\hat{M}_{B}$ in Eq.~\eqref{eq_MBstructure}, but where 
\begin{equation}
    \hat{m}_{\rm P} = \begin{pmatrix}
        \omega_{c,\mathbf{k}}-\omega_{\rm las}-i\gamma_c/2+Wn_x   & \Omega/2 + W\sqrt{n_xn_c}e^{-i\phi_{xc}}  \\ 
        \Omega/2 + W\sqrt{n_xn_c}e^{i\phi_{xc}}  & \omega_{x}-\omega_{\rm las} -i\gamma_x/2 + Wn_c
    \end{pmatrix} ~,
\end{equation}
and
\begin{equation}
    \hat{m}_{\rm HP} = 
    \begin{pmatrix}
        0            & W\sqrt{n_xn_c}e^{-i\phi_{xc}}  \\ 
        W\sqrt{n_xn_c}e^{-i\phi_{xc}} &  0 
    \end{pmatrix}~.
\end{equation}

\subsubsection{$H_4$: exciton and photon mediated saturation mechanism}
\label{sec_H4}

The interaction Hamiltonian that we consider for this mechanism is
\begin{equation}
\hat {\cal H}_{4} = (\hbar J/2) \sum_{{\bf k},{\bf k}',{\bf q}} \left(
\hat b_{{\bf k}+{\bf q}}^\dagger \hat b_{{\bf k}'-{\bf q}}^\dagger \hat a_{{\bf k}'} \hat a_{\bf k} + {\rm h.c.} \right)
\label{H_4}
\end{equation}
Assuming only this interaction term in the model, the MF equations become: 
\begin{equation}
    i\frac{\partial \psi_x}{\partial t}=\left(\omega_x-\omega_{\rm las}-i\frac{\gamma_x}{2}\right)\psi_x+\left(\frac{\Omega}{2}+J\psi_x^*\psi_c\right)\psi_c  
    \label{MFx_H4}
\end{equation}
and:
\begin{equation}
    i\frac{\partial \psi_c}{\partial t}=\left(\omega_{c,\mathbf{k_p}}-\omega_{\rm las}-i\frac{\gamma_{c}}{2}\right)\psi_c+\left(\frac{\Omega}{2}+J\psi_x\psi_c^*\right)\psi_x+F_{\rm las}.
    \label{MFc_H4}
\end{equation}
The corresponding $4\times 4$ Bogoliubov matrix $\hat{M}_{B,H4}$ has the same structure as $\hat{M}_{B}$ in Eq.~\eqref{eq_MBstructure}, but where
\begin{equation}
    \hat{m}_{\rm P} = \begin{pmatrix}
        \omega_{c,\mathbf{k}}-\omega_{\rm las}-i\gamma_c/2   & \Omega/2 + 2J\sqrt{n_xn_c}e^{i\phi_{xc}}  \\ 
        \Omega/2 + + 2J\sqrt{n_xn_c}e^{-i\phi_{xc}}  & \omega_{x}-\omega_{\rm las} -i\gamma_x/2
    \end{pmatrix} ~,
\end{equation}
and
\begin{equation}
    \hat{m}_{\rm HP} = 
    \begin{pmatrix}
        \frac{J}{2}n_x            & 0  \\ 
         0 &  \frac{J}{2}n_ce^{-2i\phi_{xc}}
    \end{pmatrix}~.
\end{equation}

The results of fitting WP3c's experimental data with these models are discussed in the next section.

\subsection{Detailed results of the comparison between all interaction models with WP3c's data}

\begin{figure}[h]
\includegraphics[width=\textwidth]{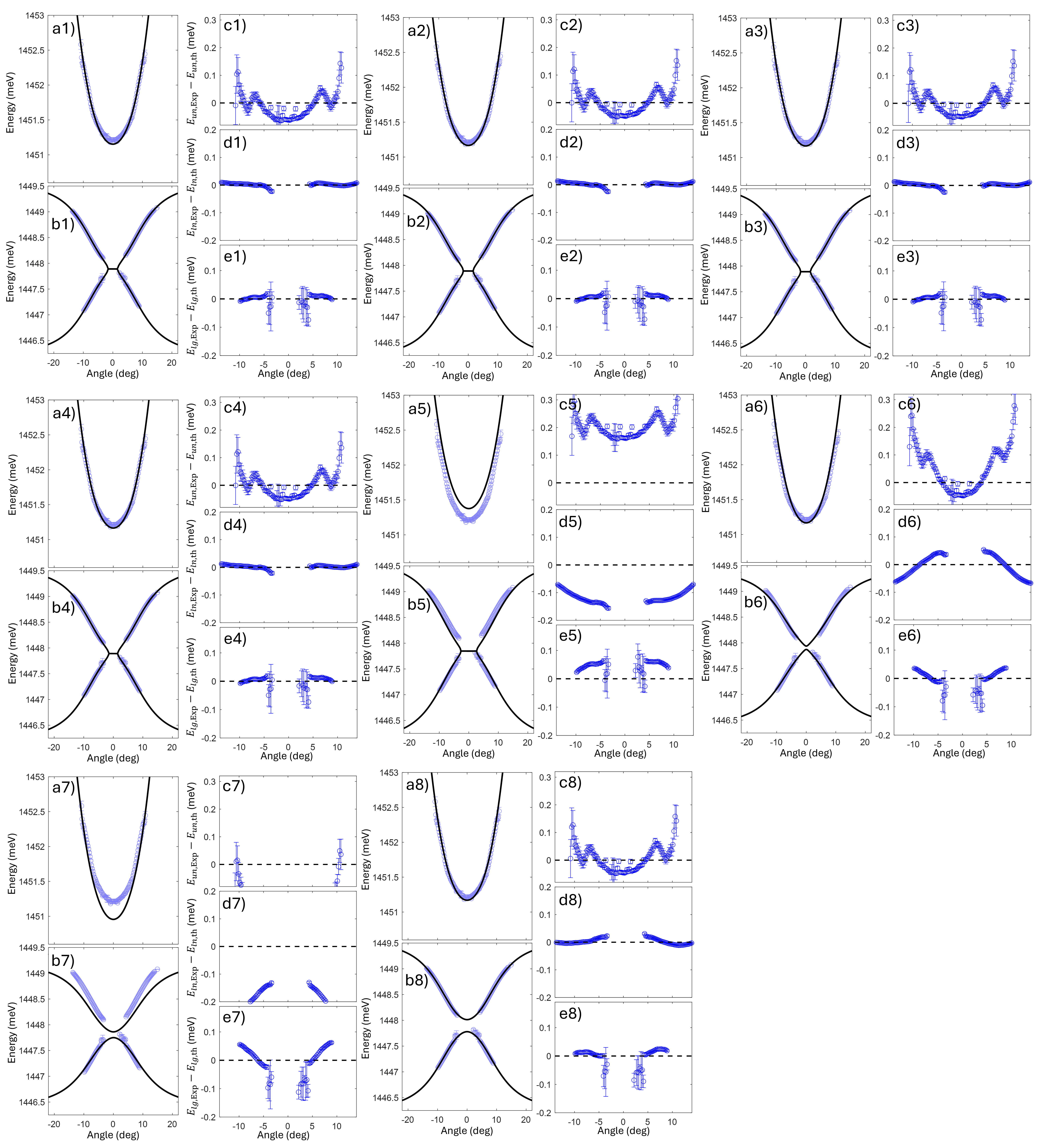}
\caption{The measured (blue symbols) polaritons Bogoliubov excitations dispersion relations for WP3c are shown in panels ($a_j$) for the upper branch and ($b_j$) for the normal and ghost lower branches. The fit with model $j$ is shown as a solid black line. The fits residuals defined as $E_{\rm Exp}-E_{\rm th}$, the difference between the measured and calculated energies are shown for the upper, normal lower and ghost lower branches in panels $c_j$) ,$d_j$) and $e_j$) respectively. $j$, the model index number is given in Table (\ref{tab:WP3_all_models})}.
\label{fig_SM4}
\end{figure}

We summarize in Table \ref{tab:WP3_all_models} the results of the fits of working point WP3c with all the models considered in this work. The corresponding fits and fit residuals are shown in Fig.\ref{fig_SM4}. Models involving the interaction terms $H_3$,$H_4$ and $H_{xx}$ all exhibit a poor agreement (lines 5,6 and 7 in Table \ref{tab:WP3_all_models}) as compared $H_{xc}$ (lines 1 in Table \ref{tab:WP3_all_models}). Adding $H_{xx}$, and/or a reservoir to $H_{xc}$ negligibly modifies the very good agreement, and results in very small relative contributions of these terms (lines 2, 3 and 4 in Table \ref{tab:WP3_all_models}). 

The combination of $H_{xx}$ and the polariton-reservoir saturating interaction described in section C.2 yields a good agreement with WP3c (line 8 \ref{tab:WP3_all_models}). However, this combination is unlikely for two reasons: (i) such a model effectively provides two independent free parameters to adjust the effective Rabi splitting on one hand (via the reservoir interaction strength $g_r^{\rm sat}n_r$), and the effective excitonic level (via $g_{xx}n_x$) on the other hand. The fact that $H_{xc}$ works even better with a single parameter makes it a much more likely explanation. (ii) The fit converges on a state with a large gap in the Bogoliubov excitations (cf. Fig.4.b8). This gap does not agree well with the few but significant low-momenta data points, and it is not consistent with the experimental situation, in which WP3c was measured: with a laser power within the bistable area of the polaritonic mean-field. Such a gap, of magnitude ($240\,\mu$eV) even larger than the laser detuning $203\,\mu$eV, means that the interaction energy in this model is more than double that of the sonic point (that occurs on the low power side of the bistable region), and hence that the model implies a laser power far away beyond the bistable region, which is not the experimental condition.

\begin{table}[h!]
    \centering
    \begin{tabular}{|c|c||c|}
    \hline
        $j$ & model & $S_{\rm fit}$ with WP3c [$\mu$eV] \\ \hline
        1&  $H_{xc}$  & 28 \\
        2&  $H_{xc}+H_{xx}$  & 27.4 \\
        3&  $H_{xc}+H_{xx}$ and $g_rn_r$  & 27.4 \\
        4&  $H_{xc}$ and $g_rn_r$  & 27.5 \\
        5&  $H_{xx}$  & 140 \\
        6&  $H_3$  & 62 \\
        7&  $H_4$  & 168 \\
        8&  $H_{xx}$ and $g_r^{\rm sat}n_r$  & 32 \\         
         \hline
    \end{tabular}
    \caption{Fit agreement of the different considered model with the data of WP3c, quantified by the fit standard error $S_{\rm fit}$. $j$ is an index referring to the corresponding panels in Fig.(\ref{fig_SM4}).}
    \label{tab:WP3_all_models}
\end{table}

\end{document}